\documentclass[a4paper,10pt]{article}

\usepackage[utf8x]{inputenc}

\usepackage{amsmath,amssymb,bm}
\usepackage{sidecap}
\usepackage{times}
\usepackage{braket}
\usepackage{ulem} 
\usepackage{graphicx,epsf}
\usepackage{float}
\usepackage[left=2.5cm,top=2.5cm,right=2.5cm,bottom=3cm,nohead,nofoot]{geometry}

\usepackage{xcolor} 

\date{}              

\begin{document}

\begin{centering}
\LARGE\textbf{Transverse polarization light scattering in tubular semiconductor nanowires}
       
\vspace{12pt}      
\normalsize\textbf{Miguel Urbaneja Torres$^{\ast}$, Anna Sitek$^{\ast,\ddagger}$, and Andrei Manolescu$^{\ast}$}

\vspace{0pt}  
\normalsize\textit{$^{\ast}$School of Science and Engineering, Reykjavik University, Menntavegur 1, IS-101 Reykjavik, Iceland}\\
\normalsize\textit{$^{\ddagger}$Department of Theoretical Physics, Faculty of Fundamental Problems of Technology, Wroclaw University of Science and Technology, 50-370 Wroclaw, Poland}\\
\normalsize\textit{e-mail: miguelt16@ru.is}\\
\end{centering}
\vspace{6pt}  

\noindent
\textbf{ABSTRACT}
\vspace{3pt}

We carry out numerical calculations of the scattering cross section of
tubular semiconductor nanocylinders in the optical range. The scattering
is investigated for the transversal incidence of light, i.e., along the 
diameter of the cylinder, with both transverse electric
and transverse magnetic polarization. These subwavelength nanostructures
support Mie resonances and, when the length of the cylinder is comparable to the
wavelength, guided modes that can overlap with the Mie modes giving rise to sharp Fano
resonances. We show that a varying internal radius affects each mode
differently, allowing for an extra degree of freedom for tuning the
spectral position of the resonant peaks.


\vspace{12pt}
\noindent
\textbf{1. INTRODUCTION}
\vspace{3pt}

The study of light-matter interaction in dielectric
nanostructures with high refractive index has recently become one of the most attracting research topics in photonics \cite{Kuznetsov1}. The interest in these structures stems from
their ability to scatter light in the optical range with, unlike their
plasmonic metallic counterparts, almost negligible losses \cite{Albella1}
while being able to support multiple resonant modes. This makes them
great candidates for building blocks of optoelectronic devices capable
of manipulating visible light at the nanoscale \cite{Kuznetsov2}. The
magnetic and electric resonances coexist within the same spectral domain 
and are highly tunable through the material and the structure geometry \cite{Vandegroep1}.
A wide range of structures, ranging from simple spherical particles \cite{Etxarri1}, and cylindrical nanowires \cite{Linyou1, vandeHaar1}, to other shapes including blocks or
pyramids \cite{Evlyukhin1} and metasurfaces \cite{Yang1}, 
have been studied and showed interesting phenomenology
with original applications. These include directional scattering
through the interference of different resonant modes \cite{Person1},
light propagation via near-field coupling between mesoscopic
scatterers \cite{Naraghi}, second and third-harmonic generation \cite{Yang2}, or Raman scattering
\cite{Caldarola1}.

Additionally, Fano resonances, which are asymmetric line-shape scattering resonances, have also attracted considerable attention in the last years \cite{Limonov1}, due to their promising applicability in optical switching and sensing. In the case of dielectric materials Fano resonances are typically obtained in dimers and oligomers. However, it has recently been shown that strong Fano resonances can also occur in simple-shaped single dielectric nanowires \cite{Abujetas1} where they are the result of the interference between a sharp guided mode and either a broad leaky mode or a Mie background.

Analytical calculations of the absorption and the scattering of light due to 
nanostructures are possible only for spherical particles (the so called Mie problem) and
for infinite cylinders \cite{Bohren}.
In our present work, we consider hollow semiconductor nanowires, i.e., nanocylinders or nanotubes, and evaluate the scattering cross section under transversal electric or magnetic (TE or TM) plane waves illumination using numerical methods. We first carry out calculations for short nanowires and show how the tubular structure affects the resonant Mie modes, in particular we study their dependence on the internal radius. Next we repeat the  simulations for longer nanowires in order to study the effects on the Fabry-Pérot guided modes and on the Fano resonances. We show that, for TE polarization, a varying internal radius strongly affects the Mie modes whereas the Fabry-Pérot guided modes are almost unaffected. These properties allow to obtain different overlapping, and thus enable the control of the Fano resonances.

\vspace{12pt}
\noindent
\textbf{2. METHODS}
\vspace{3pt}

We carry out numerical calculations of the scattering cross section
using finite element method (FEM) and perform the simulations using Comsol Multiphysics  5.4 (wave optics module)
\cite{Comsol}. In order to optimize the calculation time we take the advantage of the radial symmetry. The simulation consists of a quadrant of a sphere with 1200 nm radius in which a quarter of the nanowire, cut by two symmetry axes, is embedded in the center. The sphere consists of an outer perfectly-matched layer (PML) of 400 nm thickness and the inner space where the nanowire is located. We impose boundary conditions (perfect electric conductor [PEL] and perfect magnetic conductor [PMC]) on the facets of the quadrant for the background electromagnetic field according to the chosen polarization of the incident light. We first consider short cylinders, with the length of $L = 150$ nm, 
and the external radius of $R_{\rm ext}= 100$ nm.  Our cylinders have a tubular
shape, with a constant external radius ($R_{\rm ext}$) and the internal one ($R_{\rm int}$) varying between zero and $R_{\rm ext}$.   
We study the spectral dependence of the normalized scattering cross
section of light in the range of $\lambda = 400-800$ nm. Further, we consider
longer nanowires, i.e., of length $L = 600$ nm, and repeat the
calculations for the scattering cross section for the incident light wavelengths up to $\lambda = 1000$
nm. Finally, we obtain the spectra for the constant value of the aspect ratio $R_{\rm int}/R_{\rm ext}=1/2$, and variable external diameter ($2R_{\rm ext}$). We compare the results with the spectra obtained for the bulk nanowire. In all cases the refractive index is fixed to $n = 3.5$, which corresponds to GaP at $\lambda = 530$ nm \cite{Aspnes1}, and we assume that the nanowires are surrounded by air ($n = 1$). Since we are mostly interested in the
effects of the geometry, the absorption is neglected in all the cases. The
results are still valid and scalable for materials with negligible
dispersion in the refractive index, such as GaP in the optical range or Si
in the infrared domain.

\begin{figure} [t]
\centering
\includegraphics[scale=0.18]{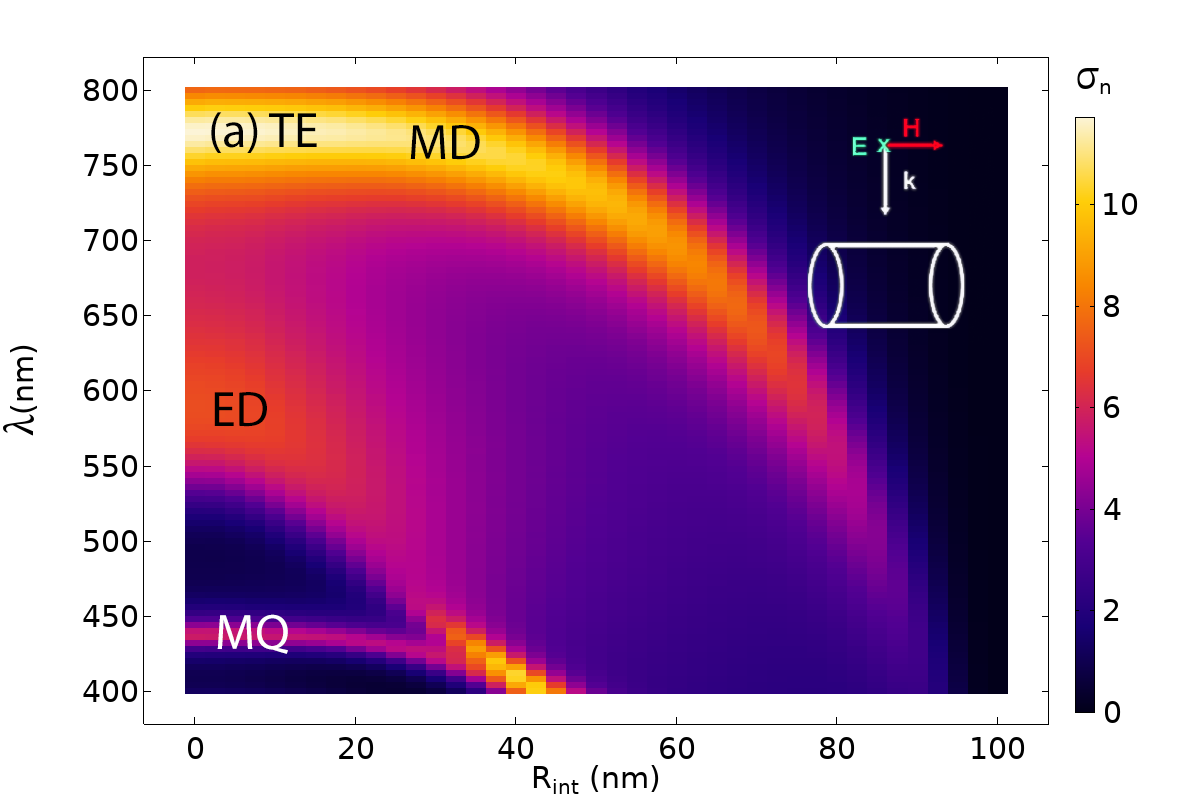}
\includegraphics[scale=0.18]{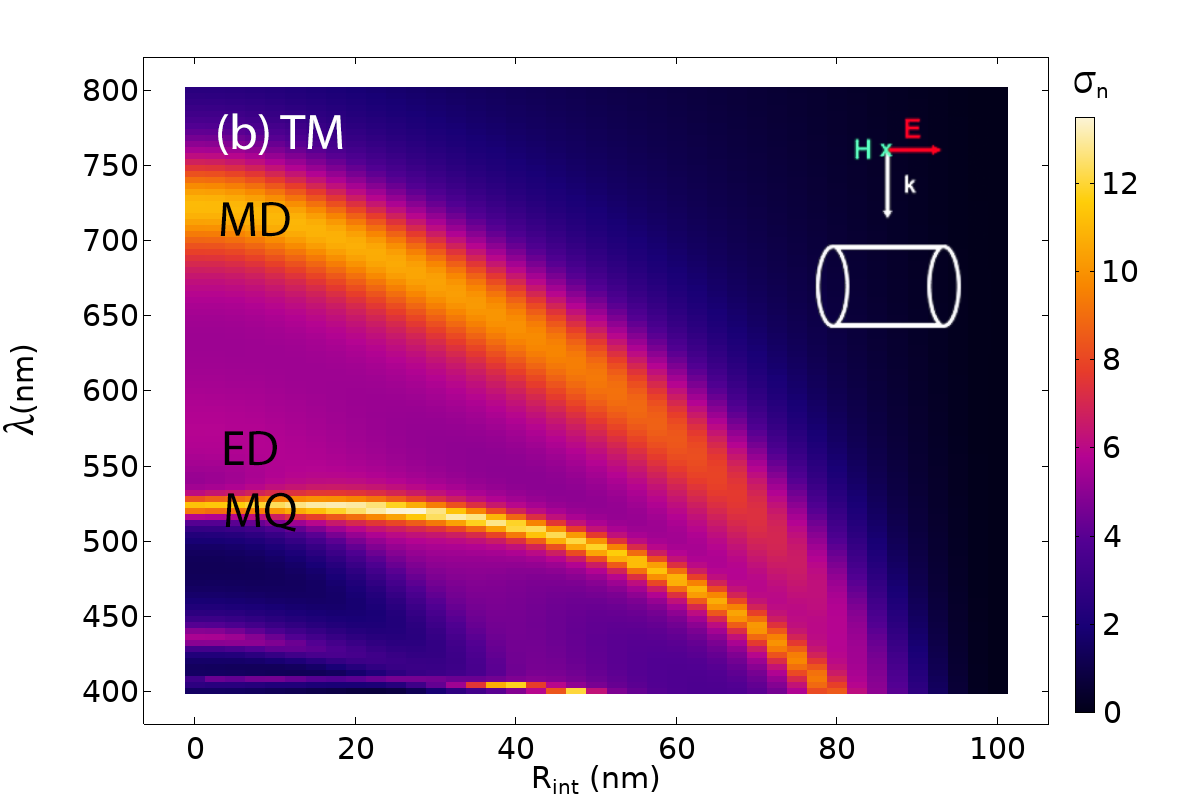}
\caption{Normalized scattering cross section spectral dependence on internal radius $R_{\rm int}$  for a short nanowire under TE polarization (a) and TM polarization (b). In both cases the nanowire length is $L = 150$ nm.}
\label{Short}
\end{figure}


\vspace{12pt}
\noindent
\textbf{3. RESULTS} 
\vspace{3pt}

\noindent
\textbf{3.1 Short nanowires}
\vspace{3pt}

First we study the scattering cross section of short cylindrical particles
in which a longitudinal hole of increasing radius is created. 
Such hollow dielectric particles, whose dimensions are smaller than the light wavelength, are
known to support strong magnetic or electric Mie resonances. A normally incident plane wave may excite
either vertical TE or TM modes \cite{vandeHaar1}.

As expected, with increasing the internal radius of the tubular
structure, the resonant scattering modes are blue-shifted, because
the thickness of the cylindrical shell is reduced, 
and thus the effective scattering
volume shrinks. However, not all the modes decay with the same rate. As shown in Fig. \ref{Short} under TE polarization, the electric dipole (ED)
shifts remarkably faster than the magnetic dipole (MD) and the magnetic
quadrupole (MQ). The MD, oriented along the axis of the cylinder, 
is almost unaffected for $R_{\rm int} < 40$ nm, since the displacement currents that create it are stronger on the outer part of the nanowire. Under TM polarization both
the ED and the MD are shifted in the same way. We also observe that the
MD mode predominates when the impinging light is TE polarized whereas under TM
polarized light the MQ becomes the dominant mode. A detailed explanation
of the behavior of these modes can be found in \cite{vandeHaar1}. \


\vspace{3pt}

\noindent
\textbf{3.1 Long nanowires}
\vspace{3pt}

As the length of the nanowires increases, the resonant Mie modes evolve until they converge to those of an infinite cylinder. These transverse
resonances may coexist with such longitudinal guided modes as Fabry-Pérot
resonances start to develop as a function of the length. This is the result of the transverse modes bouncing back and forth along the length of the nanowire. These overlapping
of transverse and longitudinal modes leads to sharp Fano resonances \cite{Abujetas1}.

\begin{figure} [H]
\centering
\includegraphics[scale=0.18]{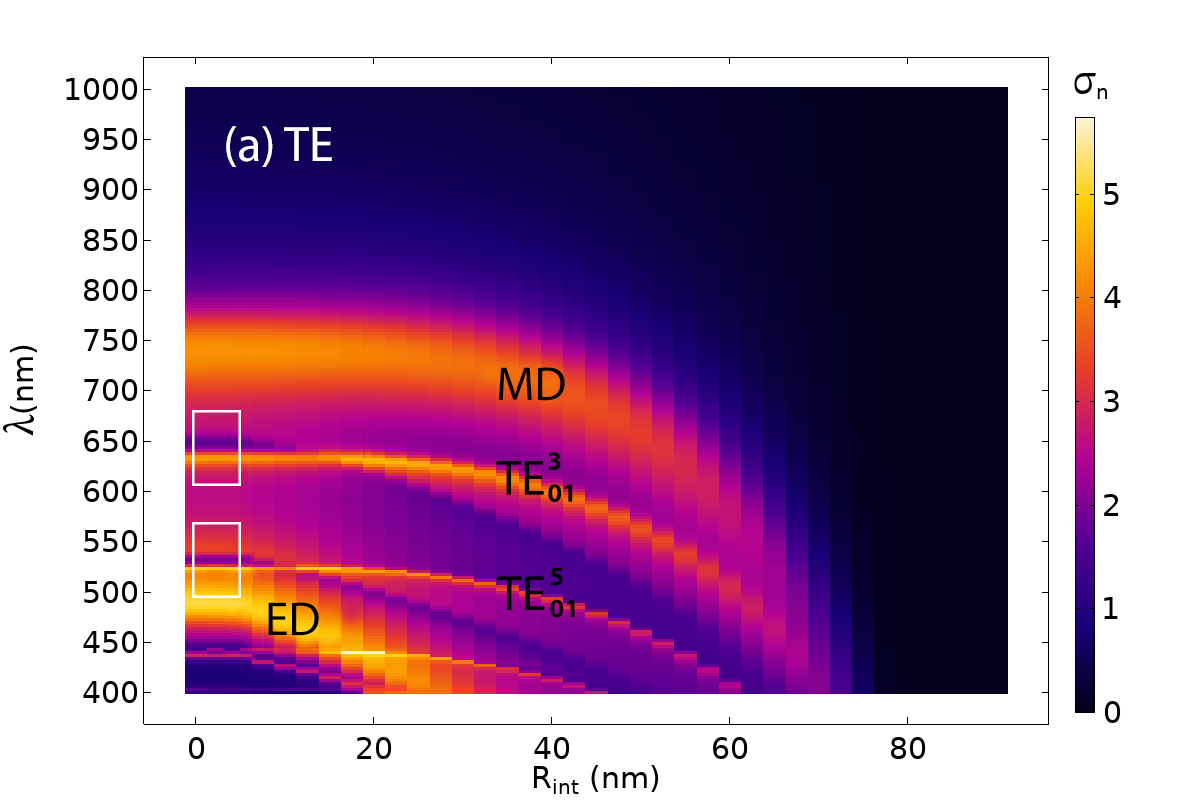}
\includegraphics[scale=0.18]{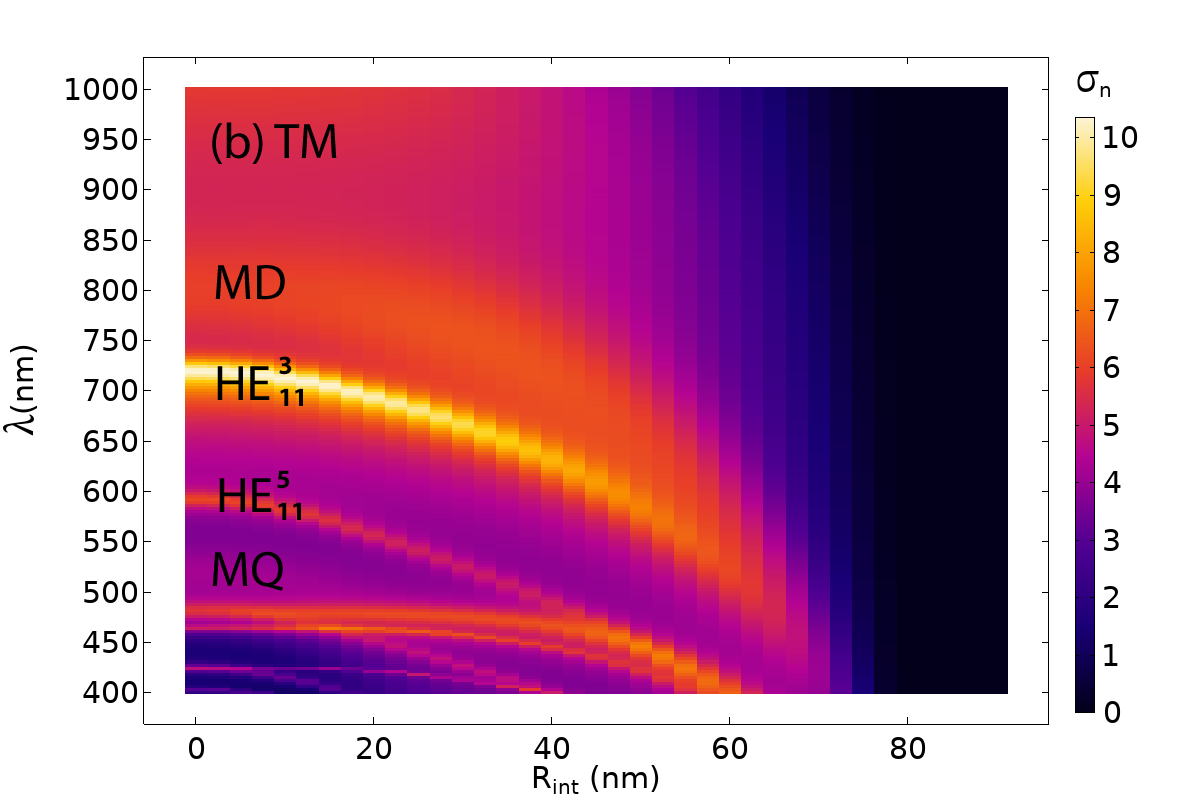}
\caption{As in Fig. \ref{Short} but for $L = 600$ nm. The rectangles indicate the ranges where the main Fano resonances occur.}
\label{Long}
\centering
\includegraphics[scale=0.18]{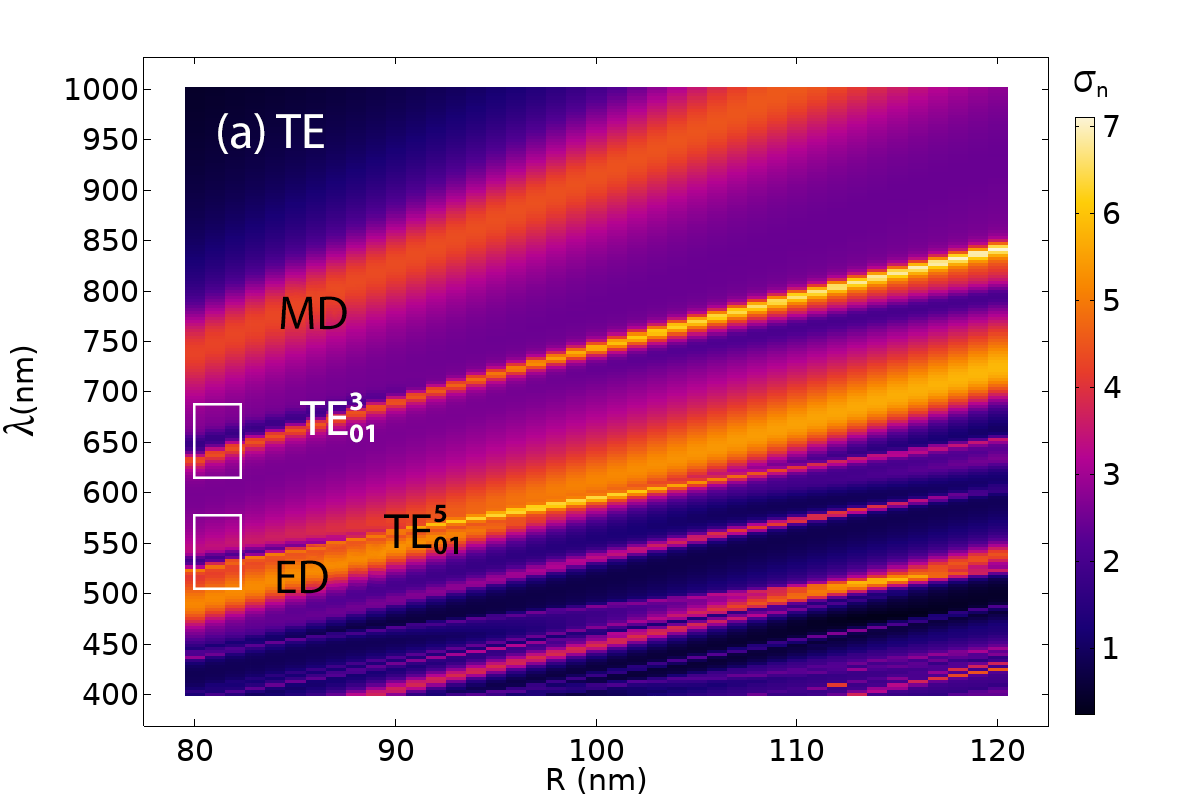}
\includegraphics[scale=0.18]{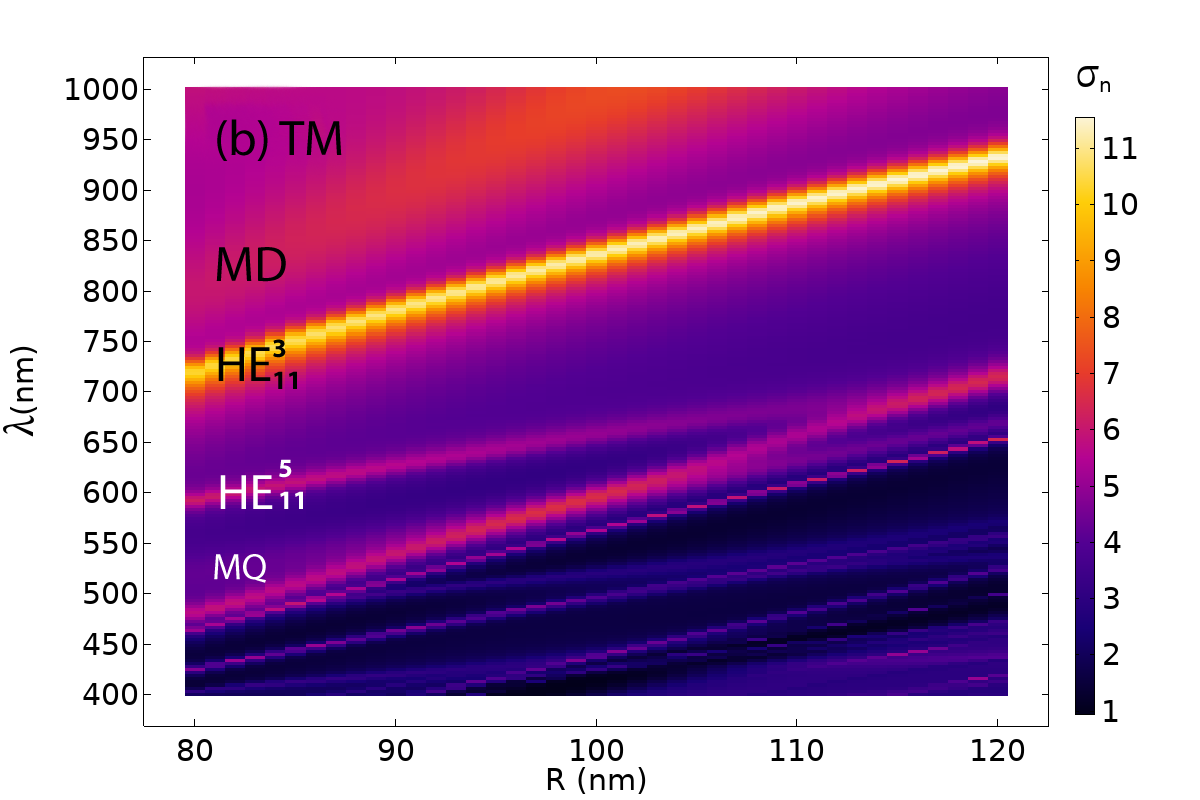}
\caption{Normalized scattering cross section spectral dependence on external radius $R_{\rm ext}$ for a (full) cylindrical nanowire under TE polarization (a) and TM polarization (b). In both cases the length is $L = 600$ nm.}
\label{LongBR}
\centering
\includegraphics[scale=0.18]{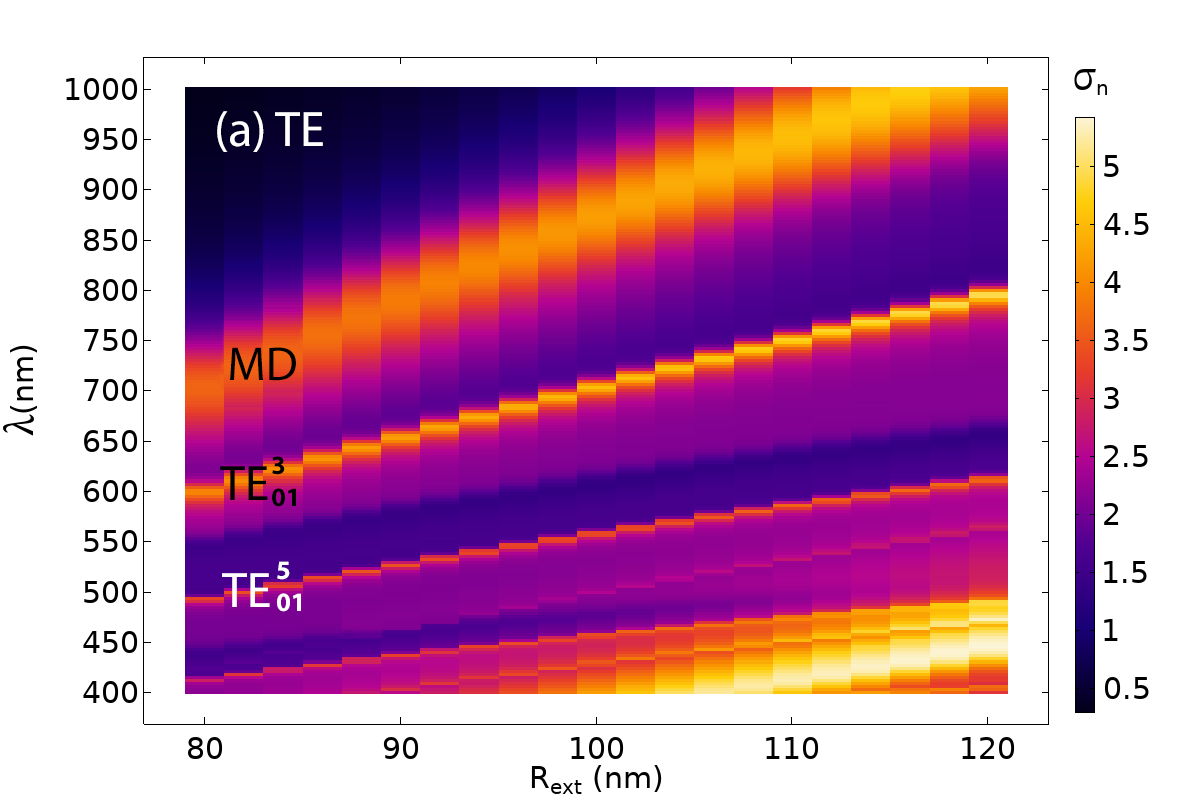}
\includegraphics[scale=0.18]{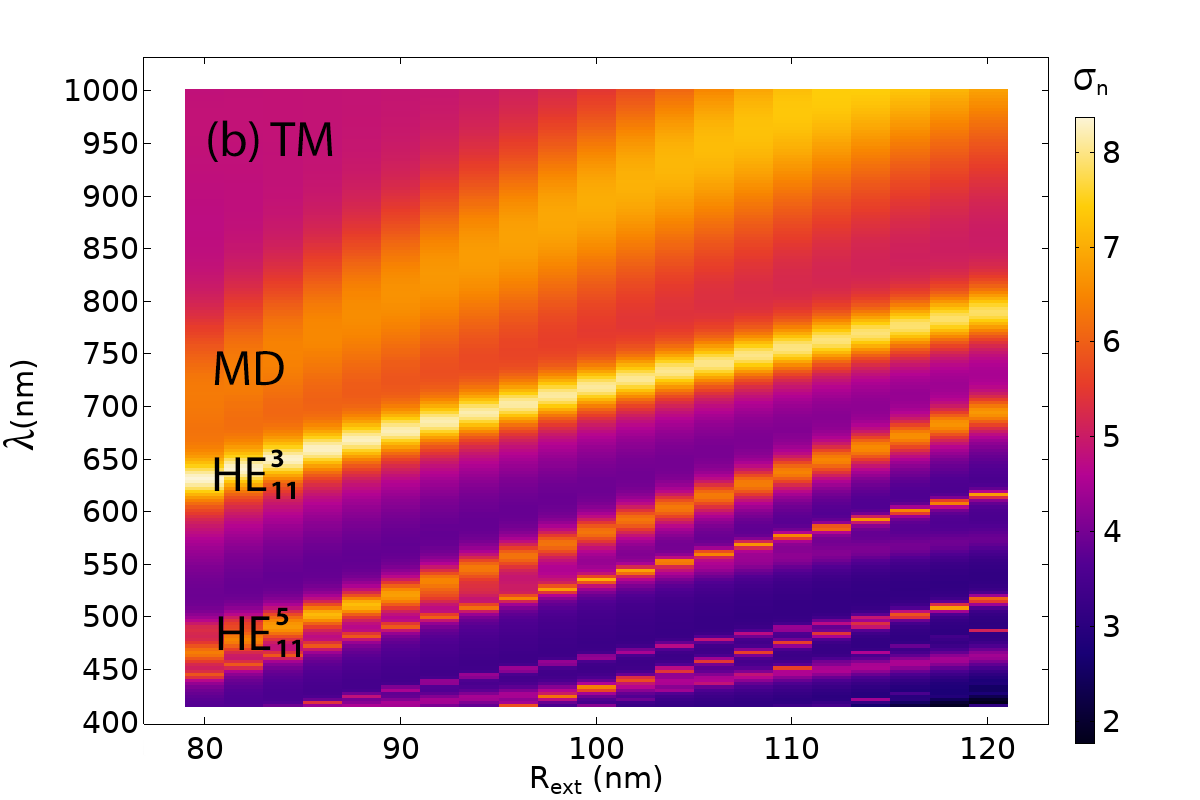}
\caption{Normalized scattering cross section spectral dependence on the external radius $R_{\rm ext}$ for a tubular nanowire with a fixed ratio $R_{\rm int} = R_{\rm ext}/2$. In both cases the length $L = 600$ nm.}
\label{LongR}
\end{figure}

Here we explore how to tailor the Fano resonances by changing the
geometry of the nanowires. The presence of the longitudinal hole
distorts the Mie transverse modes in the same way
as for the short nanowires studied in the previous section.  
However, the guided modes are affected
in a different way. As shown in Fig. \ref{Long} under TE incidence, the
guided modes (TE$_{01}$) are almost unaffected by the increasing internal radius since they are associated to the MD. This situation allows for the inversion of the symmetry of the
Fano resonances because the blue-shift of the broad ED mode is much faster. On the other hand,
under TM polarization both the Mie background and the hybrid guided modes
(HE$_{11}$) are blue-shifted at equivalent rates, and thus the crossing
of the main resonant peaks is avoided.

A detailed description of these modes 
for cylindrical nanowires, i.e., for which $R_{\rm int}=0$,
can be found in \cite{Abujetas1}.
Here we also compare the dependence of the scattering spectra on the external
radius for cylindrical (Fig. \ref{LongBR}) and tubular nanowires (Fig. \ref{LongR}) with fixed aspect ratio ($R_{\rm int} = R_{\rm ext}/2$). Now the
scattering cross section increases with the radius, and thus all the modes
are red-shifted and their spectral position depends on the size parameter
$x = \pi R_{\rm ext}^2 / \lambda$. Within the studied range of external radii ($R_{\rm ext} = 80-120$
nm) the modes are linearly red shifted with different rates. In the case
of a full nanowire exposed to TE polarized light, the increasing radius
creates a crossing of modes, which again allows for an inversion of the
Fano resonance.
However, for the tubular nanowire, the red-shifting rate of the resonant modes does not differ
much between the Mie background and the guided modes, again avoiding
any crossing in the studied range.

\vspace{12pt}
\noindent
\textbf{4. CONCLUSIONS}
\vspace{3pt}

We have studied the geometrical optical resonances and calculated the
scattering cross section of semiconductor cylindrical nanowires and the transition from bulk to 
tubular cylinders. We have shown that the tubular structure offers
an additional degree of freedom for tuning the spectral position of the
resonant modes. Since various modes are affected in different ways by the
geometrical parameters it is possible to invert Fano resonances and change
the spectral spacing between modes. This possibility is explored both in
full nanowires by changing the external radius and in tubular structures by changing the internal radius.

\vspace{12pt}
\noindent
\textbf{ACKNOWLEDGEMENTS}
\vspace{3pt}

\noindent
This work was financed by the Icelandic Research Fund, project 163438-051. 

%
%

\end{document}